\documentclass[pra,aps,twocolumn,longbibliography,nobalancelastpage,superscriptaddress]{revtex4-1}

\usepackage{amsmath}
\usepackage[utf8]{inputenc}
\usepackage[T1]{fontenc}
\usepackage{graphicx}
\usepackage{subcaption}
\usepackage{import}
\usepackage{dcolumn}
\usepackage{bm}
\usepackage{physics}
\usepackage{epstopdf}
\usepackage{amssymb}
\usepackage{mathtools}
\usepackage{amsthm}
\usepackage[normalem]{ulem}
\usepackage[toc,page]{appendix}
\usepackage{framed}
\newlength\mylength

\usepackage{xcolor}
\usepackage[colorlinks]{hyperref}
\AtBeginDocument{%
\hypersetup{
   citecolor=blue,
   linkcolor=blue,
   urlcolor=blue}}

\newcounter{protocol}

\begin{document}

\title{\textbf\title{Experimental demonstration of quantum advantage in communication complexity for Euclidean distance problem}}

\author{Verena Yacoub}
\email{verena.yacoub@lip6.fr}
\affiliation{Sorbonne Université, CNRS, LIP6, 4 place Jussieu, F-75005 Paris, France}
\author{Niraj Kumar}
\affiliation{School of Informatics, University of Edinburgh, 10 Crichton Street, Edinburgh EH8 9AB, UK} \email{While working on this project, N.K. was affiliated with University of Edinburgh. He is currently affiliated with JP Morgan Chase \& Co.}
\author{Iordanis Kerenidis}
\affiliation{Université Paris Cité, CNRS, IRIF, 8 Place Aurélie Nemours, 75013 Paris, France}
\author{Eleni Diamanti}
\affiliation{Sorbonne Université, CNRS, LIP6, 4 place Jussieu, F-75005 Paris, France}

\begin{abstract}
\normalsize When considering the complexity of communication protocols, the aim is to perform a certain task with the minimum amount of communication resources, such as time and transmitted information. The use of quantum states may lead to an exponential advantage in the use of such resources. Here, we are interested in the task of calculating the Euclidean distance between two vectors representing real data sets. It has been previously shown that it is possible to obtain an advantage for this task based on quantum fingerprinting. This protocol is defined in the simultaneous message passing model of communication complexity, where the two parties do not communicate with each other but send data to a third party, and exploits practical fingerprints generated using trains of coherent state pulses instead of highly entangled qubit states that are hard to generate for large input sizes needed to demonstrate an exponential advantage. We perform a proof-of-principle experimental demonstration of the Euclidean distance protocol using amplitude modulation techniques for encoding non-binary data sets and high-performance superconducting nanowire single-photon detectors required to increase the accessible input size. We show a quantum advantage in transmitted information surpassing the best classical protocol for an input size of $10^8$, for diverse types of data sets, including those corresponding to real grayscale images, and with reasonable precision and error bounds. Our results highlight the potential of quantum communication complexity for use in a broad set of applications. 
\end{abstract}

\maketitle

\section{Introduction}
The use of quantum resources can be advantageous for various communication tasks that are either impossible to achieve with classical resources only or require many more such resources. Prominent examples of such a quantum advantage come from the cryptographic world. For example, it is possible to achieve information-theoretic security for secret message exchange with Quantum Key Distribution~\cite{scarani2009security,diamanti2016practical} and construct computationally secure bit commitment schemes for Oblivious Transfer protocols~\cite{grilo,bartusek,diamanti24}. Beyond adversarial settings, this advantage manifests itself in the framework of quantum communication complexity, which aims to minimize the amount of communication resources shared between parties wishing to perform certain distributed tasks~\cite{dense,Zei}. These resources are the time it takes to transmit the information required to perform the task or the amount of this transmitted information. Early studies in this field highlighted the role of entanglement in protocols based on teleportation or (super) dense coding to reduce the communication complexity~\cite{Ent,Ent2}. Later works used the quantum switch or the superposition of communication directions to show an exponential advantage by exploiting the indefinite causal order of operations~\cite{superposition,switch}. The quantum Zeno effect, namely the suppression of unitary evolution by frequent measurements, was also shown to lead to an advantage in classical complexity problems~\cite{Zeno}, and it was furthermore demonstrated that quantum systems have an inherent advantage in generating randomness in communication scenarios~\cite{seefledguy}.

In addition to the above examples, a major task exemplifying quantum communication complexity and that is central to our work is quantum fingerprinting~\cite{Harry}. The concept underpinning fingerprinting is that to perform a distributed task, it is sufficient to use only part of the input data. The size of the fingerprint is then bounded by the maximum error allowed for the solution. The advantage offered by quantum fingerprinting has long been established. It is known in particular that it eliminates the need for pre-shared randomness, which reduces exponentially the complexity compared to classical protocols. The experimental demonstration of such an advantage was, however, hindered by the nature of quantum fingerprints, which are multiqubit entangled states of large dimension. This difficulty was overcome by the mapping introduced in~\cite{map}, allowing for the use of coherent states, which are readily available experimental resources, together with linear optics and single-photon detection, to perform quantum communication complexity tasks in practice. This unlocked a series of experiments in the field, including the demonstration of quantum advantage in transmitted information for the task of Sampling Matching, which was derived from Hidden Matching in the coherent state fingerprint framework~\cite{Sampling}. It was also possible to show an advantage for the Equality task, first in transmitted information~\cite{Xu} and then, in principle, in both transmitted information and communication time~\cite{effprint}, leveraging an efficient protocol involving channel multiplexing and simultaneous detection techniques, in the simultaneous message passing (SMP) model of communication complexity, as originally proposed in~\cite{Ar}.
 
Here, we are interested in the communication task defined in~\cite{Niraj}, which aims to calculate the Euclidean distance (ED) between two real vectors with a precision similar to the best classical protocol but with an exponential advantage in transmitted information due to the use of quantum fingerprints. This task finds its interest as a basic function in machine learning kernels, the ED protocol
is directly relevant to distributed kernel-based machine
learning~\cite{Scholkopf2002, Suresh2017}, where estimating
the Euclidean distance is a primitive operation for computing
similarity kernels such as the Gaussian kernel between
feature vectors held by separate parties. Like for the Equality task, we focus on the SMP model, where the two parties are not allowed to communicate with each other but instead send their data to a referee, who performs suitable operations and evaluates the outcome of the task. The data is communicated only once, without any further classical or quantum communication. In our coherent state fingerprint framework, the non-binary vector information is encoded in the amplitude of the coherent states. The advantage relies on the randomness of the time of arrival of the states and the statistics obtained by the single-photon detectors used to calculate the Euclidean distance, without needing shared randomness as in the classical case. We experimentally verify the obtained quantum advantage in transmitted information for real data sets and realistic error bounds.

\section{Results}
\noindent\textbf{Overview.} Compared to prior coherent-state fingerprinting
experiments~\cite{Xu,effprint}, the present work introduces
three technical advances that together enable the first
experimental demonstration of a quantum communication
complexity protocol beyond binary data. First, we implement
\emph{amplitude-modulated encoding} over a 10-level alphabet,
mapping non-binary data directly onto the amplitudes of weak
coherent pulses rather than restricting to binary phase
modulation as in the Equality protocol. Second, we operate
at mean photon numbers per pulse as low as
$\bar{\mu}_{\mathrm{pulse}} \sim 10^{-5}$ at $n=10^8$,
which requires superconducting nanowire single-photon
detectors (SNSPDs) with quantum efficiency $\eta > 84\%$
and dark-count probability $p_d = 10^{-7}$; standard
avalanche photodiodes are insufficient at this regime.
Third, we implement \emph{real-time visibility-gated
acquisition} without active phase locking: the interferometric
phase is tracked continuously via the click statistics of
both detectors, and data acquisition is gated to periods
where the visibility exceeds the threshold $\mathcal{V}_{\min}$
required by the error bound, mitigating phase fluctuations
without the complexity of a feedforward stabilisation loop. \\

\noindent\textbf{Communication model and Euclidean distance problem.}
Our protocol is defined in the  SMP
model of communication complexity. In this model, two parties, Alice and Bob,
receive inputs $x$ and $y$ from input sets $\mathcal{X}$ and $\mathcal{Y}$
respectively. Each party sends a single message, based solely on their own
input, to a referee whose goal is to compute a function
$f(x,y):\mathcal{X}\times\mathcal{Y}\to\{0,1\}$ with failure probability at
most $\delta$. The underlying motivation for performing fingerprinting in this
model is to minimize the communication resources required, by sending only a
compact representation of the input rather than the input itself. In
communication complexity, an \emph{efficient} protocol minimizes two
resources: the \emph{communication time}, defined as the number of time units
required to transmit the messages, where one time unit corresponds to
sending a single bit, and the \emph{transmitted information}, defined as the
number of bits of information about the inputs actually conveyed through the
channel.

In classical communication, fingerprinting is implemented using a shared
random key between Alice and Bob that specifies which bits of their inputs to
send to the referee. The size of the resulting fingerprint is generally bounded
by the tolerable error margin on the referee's computation~\cite{Harry}.
Quantum fingerprinting offers a significant advantage: shared randomness can
be eliminated and replaced by local private coins at each party, while still
achieving an exponentially smaller fingerprint size. In~\cite{Ar}, Arrazola
and L\"{u}tkenhaus proposed a quantum fingerprinting scheme for the Equality
problem using only weak coherent pulses, building on the mapping of~\cite{map}
between quantum communication protocols based on multi-qubit pure states and
those based on coherent states, linear optics, and single-photon detection.

In this framework, coherent states are generated by applying the displacement
operator to the vacuum. The fingerprint sent by Alice is a train of $n$
coherent pulses,
\begin{equation}
    \ket{\alpha_x}=\hat{D}_x (\alpha)\ket{0}= \bigotimes_{j=1}^n
    \ket{x_j\alpha}_j,
    \label{eq:fingerprint}
\end{equation}
where $\hat{D}_x (\alpha)= e^{\alpha \hat{a}_x^\dagger-\alpha^* \hat{a}_x}$,
and $\hat{a}_x = \sum_{j=1}^n x_j \hat{b}_j$ is the annihilation operator of
the fingerprint mode, with $\hat{b}_j$ the photon annihilation operator at the
$j_{\text{th}}$ time bin. The amplitude $x_j$ of the $j_{\text{th}}$ pulse
encodes the $j_{\text{th}}$ component of Alice's input vector. The total mean
photon number of the fingerprint is $\mu = \sum_{j=1}^n |x_j \alpha|^2 =
|\alpha|^2$, which is independent of the input size $n$ for unit-norm input
vectors. Bob prepares his fingerprint $\ket{\alpha_y}$ analogously.

The private coins in the scheme of~\cite{Ar} arise from the probabilistic
arrival times of photons within the weak coherent pulse trains; see
Eq.~(\ref{eq:fingerprint}). These arrival times follow a hypergeometric
distribution~\cite{Xu}. As a result, the transmitted information is
exponentially reduced from $O(\sqrt{n})$ in the best classical protocol to
$\Omega(\log n)$, demonstrating an exponential gap in communication
complexity. However, since the photon arrival times are unknown in advance,
the entire $n$-pulse sequence must be transmitted to complete the protocol,
yielding a communication time of $O(n)$. In~\cite{Ar,Xu}, the mean photon
number $\mu$ was optimized to match the error bound of the best classical
protocol while achieving an exponential advantage in transmitted information.
This protocol was experimentally demonstrated for binary data in~\cite{Xu}.

\begin{figure}[ht!]
    \centering
    \includegraphics[scale=0.5]{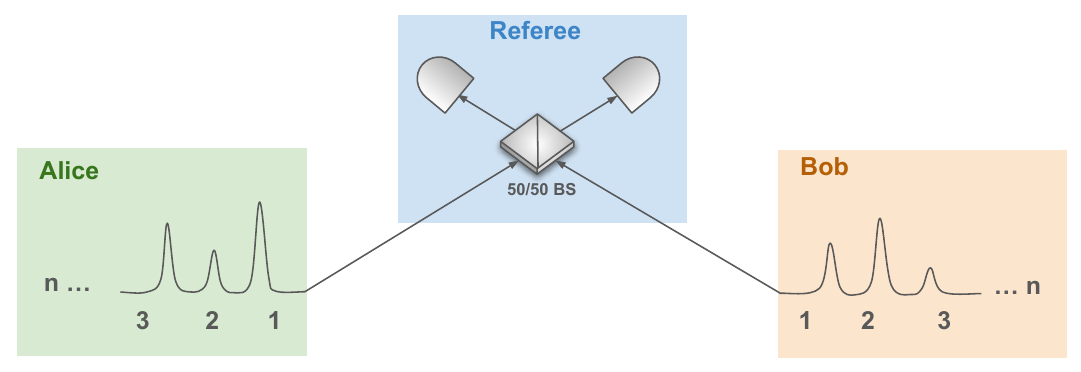}
    \caption{General scheme of the Euclidean distance protocol in the
    simultaneous message passing model. The real valued information vector is encoded
    on the amplitudes of the coherent pulses constituting the fingerprints.}
    \label{fig:sch}
\end{figure}

The Euclidean distance (ED) problem, introduced theoretically in~\cite{Niraj},
extends the coherent-state fingerprinting framework beyond binary strings to
real-valued vectors, thereby opening the door to applications in machine
learning and distributed data comparison. The general scheme is illustrated
in Fig.~\ref{fig:sch}.   The best known classical SMP protocol for the ED problem is the Equality protocol
of~\cite{Harry} applied after error-correcting encoding, achieving
a transmitted information of $4k\sqrt{n}$ bits. No classical
protocol specifically designed for ED and outperforming this bound
is known, because ED is at least as hard as Equality: any classical
ED protocol immediately solves Equality (since $\|x-y\|_2^2 = 0$
if and only if $x = y$ for unit vectors), and hence inherits the
$\Omega(\sqrt{n})$ classical lower bound proven by Newman and
Szegedy~\cite{NewmanSzegedy} for Equality in the SMP model.

Crucially, both the lower bound and the best known classical
protocol apply to the \emph{private-coin} SMP model, in which
Alice and Bob have access only to their own local randomness and
no shared random key. This is the relevant setting here: with
shared randomness, Equality (and hence ED) can be solved with
$O(1)$ communication~\cite{NewmanSzegedy}, making the problem
trivial. Our quantum protocol operates in the private-coin model, the probabilistic photon arrival times within the weak coherent
pulse trains serve as the local private coins for Alice and Bob,
as established in~\cite{Ar}, and it is precisely in this model
that the exponential quantum advantage in transmitted information
is demonstrated.

The task is for the referee to estimate the squared Euclidean distance between
the real-valued input vectors $x, y \in \mathbb{R}^n$ of Alice and Bob:
$\|x-y\|^2_2 = \sum_{j=1}^n (x_j - y_j)^2$, which is equivalently expressed
as a function of their inner product. The protocol proceeds as follows:
\begin{itemize}
    \item Alice and Bob each encode their input vectors $X$ and $Y$ on the
    amplitudes of the coherent pulses constituting their respective
    fingerprints;
    \item They simultaneously send their modulated fingerprints to the referee,
    where the pulses interfere on a 50/50 beam splitter (BS), as shown in
    Fig.~\ref{fig:sch};
    \item The referee estimates the Euclidean distance from the detection
    statistics accumulated by two single-photon detectors, $D_0$ and $D_1$,
    placed at the two output ports of the BS.
\end{itemize}

Neglecting experimental imperfections, the output state after the 50/50 BS
at the $j_{\text{th}}$ time bin is
\begin{equation}
    \ket{\frac{(x_j+y_j)}{\sqrt{2}} \alpha}_{j,D_0} \otimes
    \ket{\frac{(x_j-y_j)}{\sqrt{2}} \alpha}_{j,D_1},
\end{equation}
and the click probability at each detector reads
\begin{equation}
    p_j^{D_0} = 1- e^{\frac{-\mu (x_j + y_j)^2}{2}}, \quad
    p_j^{D_1} = 1- e^{\frac{-\mu (x_j - y_j)^2}{2}}.
\end{equation}
As shown in~\cite{Niraj}, expanding these expressions to first order in $\mu$
(valid in the weak-pulse regime) and incorporating two standard experimental
imperfections, namely the beam-splitter interferometer visibility $\nu \in
[0,1]$, which quantifies the interference contrast, and the detector
dark-count probability per time bin $p_d$, the click probabilities become
\begin{widetext}
\begin{equation}
 p_j^{D_0} \approx \frac{\mu}{2} \big( \nu(x_j+y_j)^2+(1-\nu)
    (x_j-y_j)^2+ 2\sqrt{\nu (1-\nu)} (x_j^2-y_j^2) \big) + p_d,    
\end{equation}
   
\begin{equation}
    p_j^{D_1} \approx \frac{\mu}{2} \big( \nu(x_j-y_j)^2+(1-\nu)
    (x_j+y_j)^2+ 2\sqrt{\nu (1-\nu)} (x_j^2-y_j^2) \big) + p_d.
\end{equation}
\end{widetext}
The Euclidean distance estimator is then constructed from the signed
click-count difference as
\begin{equation}
  \tilde{E} = 2 - \frac{1}{\mu(2\nu-1)} \mathbb{E} \left[ \sum_{j=1}^{n}
  \left( Z_j^{D_0} - Z_j^{D_1} \right) \right],
\end{equation}
where $Z^{D_{0,1}}\in\{0,1\}$ are Bernoulli random variables indicating
the presence or absence of a detection click at each detector and time bin.

The optimal value of $\mu$, for given visibility $\nu$,
dark-count rate $p_d$, and target precision $(\epsilon, \delta)$,
is determined by requiring that the estimator $\tilde{E}$
satisfies the \emph{additive} error bound
\begin{equation}
    \bigl|\tilde{E} - \|x-y\|_2^2\bigr| \leq \epsilon
    \label{eq:additive_bound}
\end{equation}
with probability at least $1-\delta$. Since
$\tilde{E} = 2 - \Delta Z/[\mu(2\nu-1)]$, this is equivalent
to requiring an absolute deviation bound on the signed
click-count difference:
\begin{equation}
    \bigl|\Delta Z - \mathbb{E}[\Delta Z]\bigr|
    \leq \epsilon\,\mu(2\nu-1).
    \label{eq:abs_deviation}
\end{equation}
We explicitly express the analytical upper bound on the left-hand side 
by the following Chernoff-type inequality~\cite{mitzenmacher2017}:
\begin{widetext}
\begin{equation}\label{eq:chernoff}
    \operatorname{Pr}\!\left(|\Delta Z-\mathbb{E}[\Delta Z]|
    \geqslant \epsilon\,\mu(2\nu-1)\right)
    \leqslant
    2\left(\frac{e^{r\epsilon}}{(1+r\epsilon)^{1+r\epsilon}}
    \right)^{2\mu(2\nu-1)} = \delta,
    \quad
    r = \frac{2\nu-1}{\nu+\dfrac{n-2\mu}{2\mu}\,p_d},
\end{equation}
\end{widetext}
Noting that this analytical expression we provide here was discussed in~\cite{Niraj} but never shown explicitly.

\begin{figure}[ht!]
    \centering
    \includegraphics[scale=0.35]{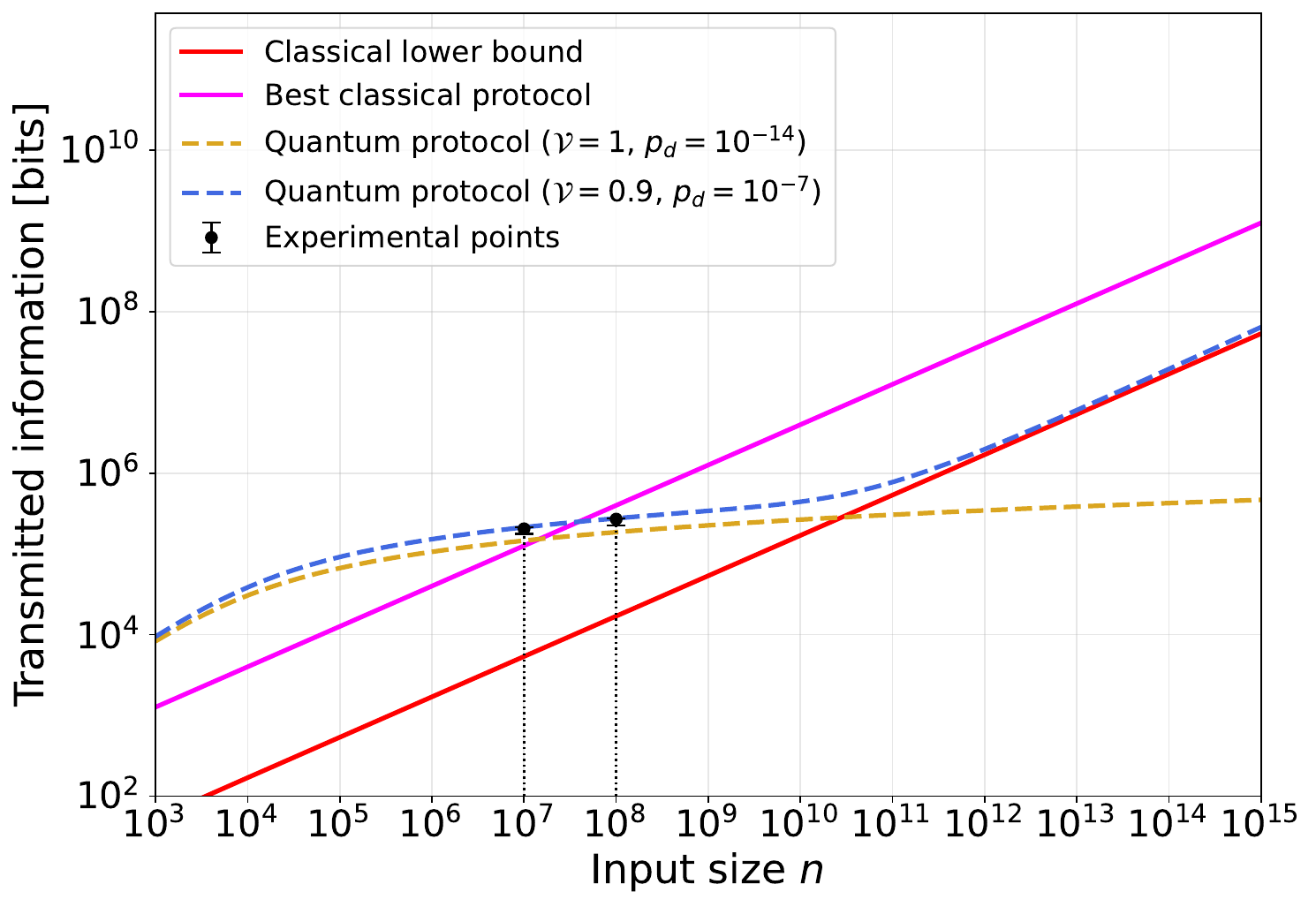}
    \caption{   Log-log plot of the transmitted information as a
function of input size $n$ for solving the ED problem with
additive accuracy $\epsilon=0.1$ and failure probability
$\delta \leq 10^{-6}$.   We compare the ideal
    ($\mathcal{V}=1$, $p_d \approx 10^{-14}$) and realistic ($\mathcal{V}=0.9$,
    $p_d \approx 10^{-7}$) ED protocol against the classical lower bound
    and the best known classical protocol for Equality. Experimental points
    are shown with black symbols. Quantification of the the quantum advantage is in Section~\ref{sec:postpro} and the detailed calculations in Appendix~\ref{app:error}.}
    \label{fig:EDplot}
\end{figure}

Note that at $\nu=1$ and $p_d=0$ one has $r=1$ and
Eq.~(\ref{eq:chernoff}) reduces to the standard multiplicative
Chernoff bound~\cite{mitzenmacher2017}. For $\nu < 1$, the factor
$r < 1$ reflects the reduced signal-to-noise ratio due to imperfect
interferometer visibility, and both the numerator and the exponent
of the denominator must be scaled consistently by $r$ for the bound
to remain valid. In practice, rather than inverting
Eq.~(\ref{eq:chernoff}) analytically, $\mu^*$ is determined
numerically using a tighter Gaussian approximation of the error
probability, as detailed in Protocol~\ref{prot:pre}. For constant
$\epsilon$ and $\delta$, the communication time scales as $O(n)$
while the transmitted information scales as $O(\mu \log n)$. In
Fig.~\ref{fig:EDplot} we show the transmitted information as a function of
input size $n$ for the ED protocol under both ideal ($\mathcal{V}=1$,
$p_d \approx 10^{-14}$) and realistic ($\mathcal{V}=0.9$, $p_d \approx 10^{-7}$)
conditions, where $\mathcal{V}=(\nu+1)/2$ and compare it against the classical lower bound derived
in~\cite{Niraj} and the best known classical protocol for Equality~\cite{Ar,
Niraj}.

We now describe our experimental implementation of the ED protocol and the
demonstration of a practical quantum advantage in communication complexity
for realistic data sets.

\section{Methods}

\begin{figure*}[t]
    \centering
     \includegraphics[scale=0.18]{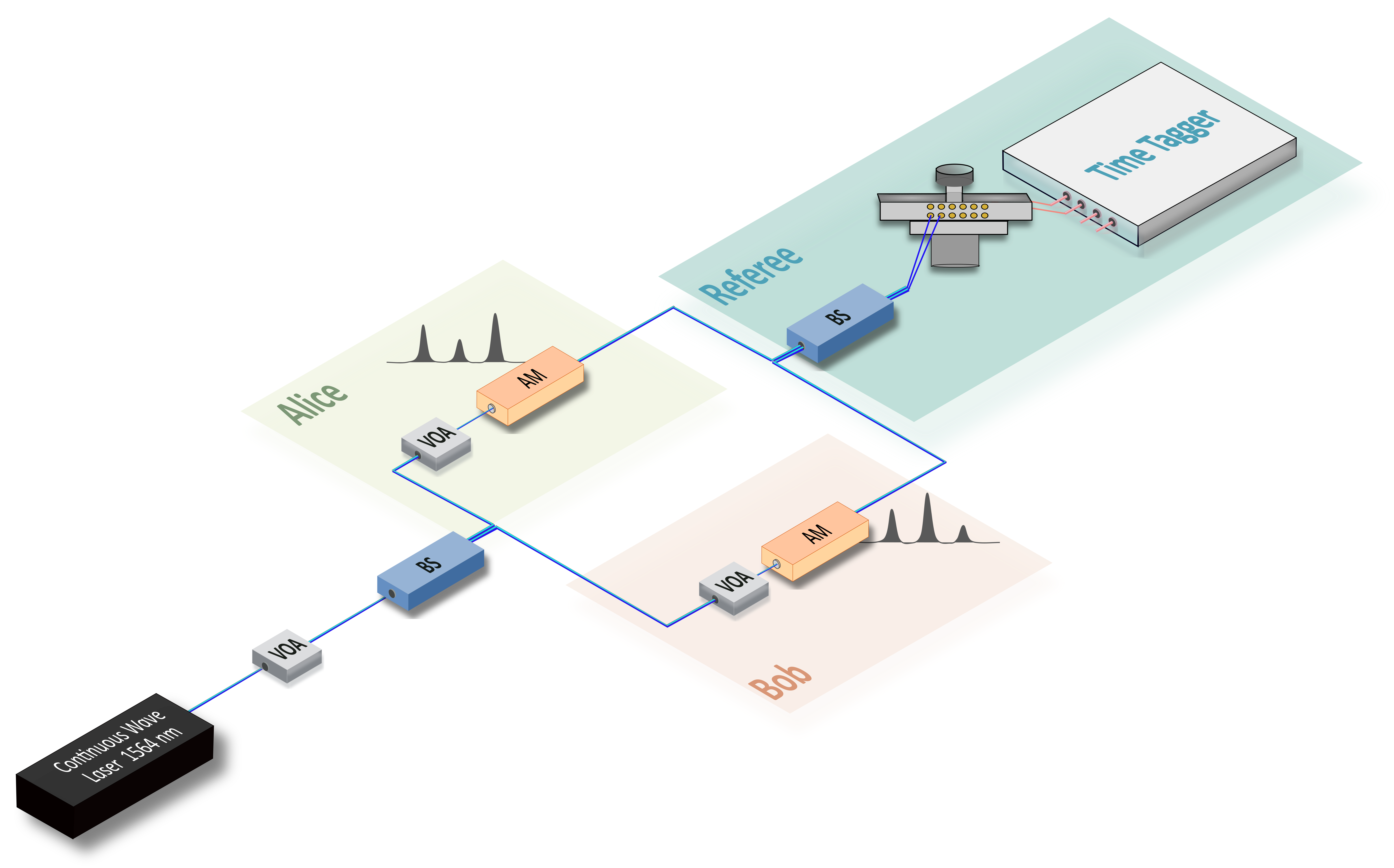}
    \caption{Experimental setup for the implementation of the Euclidean
    distance protocol.}
    \label{fig:setup}
\end{figure*}

\noindent\textbf{Setup.}
The experimental setup is shown in Fig.~\ref{fig:setup}. In our
proof-of-principle demonstration, a single continuous laser source operating at 1564~nm
is split by a 50/50 BS and directed to Alice and Bob, providing them with
phase-coherent light from which they generate their respective fingerprints.
We stress that there is no communication between Alice and Bob after this
initial beam-splitting step. Each party attenuates their signal using a
variable optical attenuator (VOA) set according to the target value of $\mu$,
and then modulates the pulse amplitudes using an electro-optic amplitude
modulator (Exail) to encode the components of their input vectors. The
modulation signals consist of 16~ns synchronous electrical pulses at a
repetition rate of 5~MHz, with amplitudes controlled by an arbitrary waveform
generator (AWG) ({Keysight M3300A}). The available modulation range supports up to 10
discrete amplitude levels, enabling the encoding of real-valued data drawn
from a 10-symbol alphabet. 

We note that while the theoretical protocol of~\cite{Niraj}
handles arbitrary real-valued inputs $x_j \in \mathbb{R}$,
our implementation encodes the input vectors using 10 discrete
amplitude levels, mapping pixel values $\{0,1,\ldots,9\}$ to
amplitudes $\{0, 1/9, 2/9, \ldots, 1\}$. The discretized vectors
are then $\ell_2$-normalised before transmission, ensuring the unit-norm
$\|\tilde{x}\|_2 = \|\tilde{y}\|_2 = 1$ as required by the
protocol. The effect of discretization on the Euclidean distance
estimator $\tilde{E}$ can be bounded by expanding:
\begin{equation*}
    |\tilde{E}_q - \tilde{E}|
    \leq 2|\langle x - y,\, d^x - d^y\rangle|
    + \|d^x - d^y\|_2^2,
\end{equation*}
where $d^x = \tilde{x} - x$ and $d^y = \tilde{y} - y$ are
the normalised quantisation error vectors. By the Cauchy-Schwarz inequality followed by the triangle
inequality and the unit-norm constraint $\|x\|_2 = \|y\|_2 = 1$:
\begin{equation*}
    |\langle x-y,\, d^x-d^y\rangle|
    \leq \|x-y\|_2\,\|d^x-d^y\|_2
    \leq 2\,\bigl(\|d^x\|_2 + \|d^y\|_2\bigr),
\end{equation*}
so the dominant term scales
as $\|d^x\|_2 + \|d^y\|_2$, which depends on
how the quantisation errors accumulate across the $n$
components. In the worst case discretization error \footnote{the worst case error is usually considered as lying on half of the step size, in our case it is $1/18$} this bound exceeds $\epsilon$,
so the negligibility of discretization error cannot be guaranteed
analytically in general. Instead, it is directly verified
experimentally: the fact that all estimated distances fall
within the $\pm\epsilon = \pm 0.1$ error bound in all
protocol runs, as shown in Figs.~\ref{fig:hist}
and~\ref{fig:hist2}, confirms that this discretization does not
affect the validity of the error bound for our specific
data sets. The 10-level encoding already represents a
significant departure from the binary setting of all
previous experimental implementations~\cite{Xu}. 
  
The modulated weak coherent pulses from Alice and Bob are directed to a second
50/50 BS held by the referee, where they interfere. The two output ports are
monitored by SNSPDs (ID Quantique), chosen for their high efficiency and low dark-count rates.
The dark-count probability for our 5~MHz pulse rate was $p_d = 10^{-7}$ for
both detectors, and their quantum efficiencies were $84\%$ and $85\%$ for
$D_0$ and $D_1$ respectively.

\vspace{0.5em}
\noindent\textbf{Pre-processing.}
Prior to each protocol run, the mean photon number $\mu$ and the
interferometer visibility $\nu$ are calibrated following
Protocol~\ref{prot:pre}. Detection events are recorded by a time tagger
(Swabian) and relayed in real time to the AWG. Rather than waiting for the
full $n$-pulse transmission, $\mu$ is estimated from a short calibration
sequence of maximum-amplitude pulses, from which the click rate is used to
infer the mean photon number per pulse. Following our theoretical model, the mean photon number per pulse $\bar{\mu}_{\mathrm{pulse}}$ is tuned such that the total mean photon number over the full fingerprint of $n$ pulses, $\mu = \bar{\mu}_{\mathrm{pulse}}
\cdot n$, lies roughly in the range $\mu \in [5000, 7500]$.
Since $\mu$ is constant and almost independent of $n$ by construction~\cite{map},
the mean photon number per pulse decreases with input size, reaching
$6 \times 10^{-4}$ at $n=10^7$ and $6 \times 10^{-5}$ at $n=10^8$. Operating
reliably at such low photon numbers requires detectors with both high quantum
efficiency and low dark-count rates, conditions met by our SNSPDs. This
allowed us to successfully run the protocol up to $n=10^8$; beyond this input
size, the real-time feedback loop could not be sustained at our pulse
repetition rate.\\

\begin{figure}[ht!]
    \centering
    \includegraphics[scale=0.3]{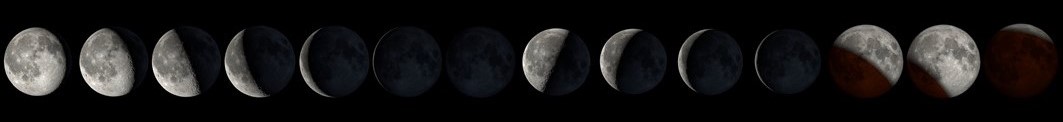}
    \includegraphics[scale=0.3]{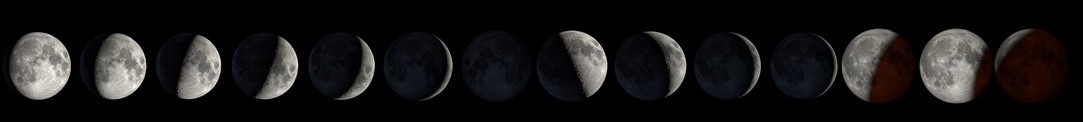}
    \caption{Grayscale images used as Alice's (upper row) and Bob's (lower
    row) input data sets in the real-data experiment.}
    \label{fig:moons}
\end{figure}

\begin{figure}[ht!]
    \begin{subfigure}[b]{0.45\textwidth}
    \includegraphics[width=0.8\linewidth]{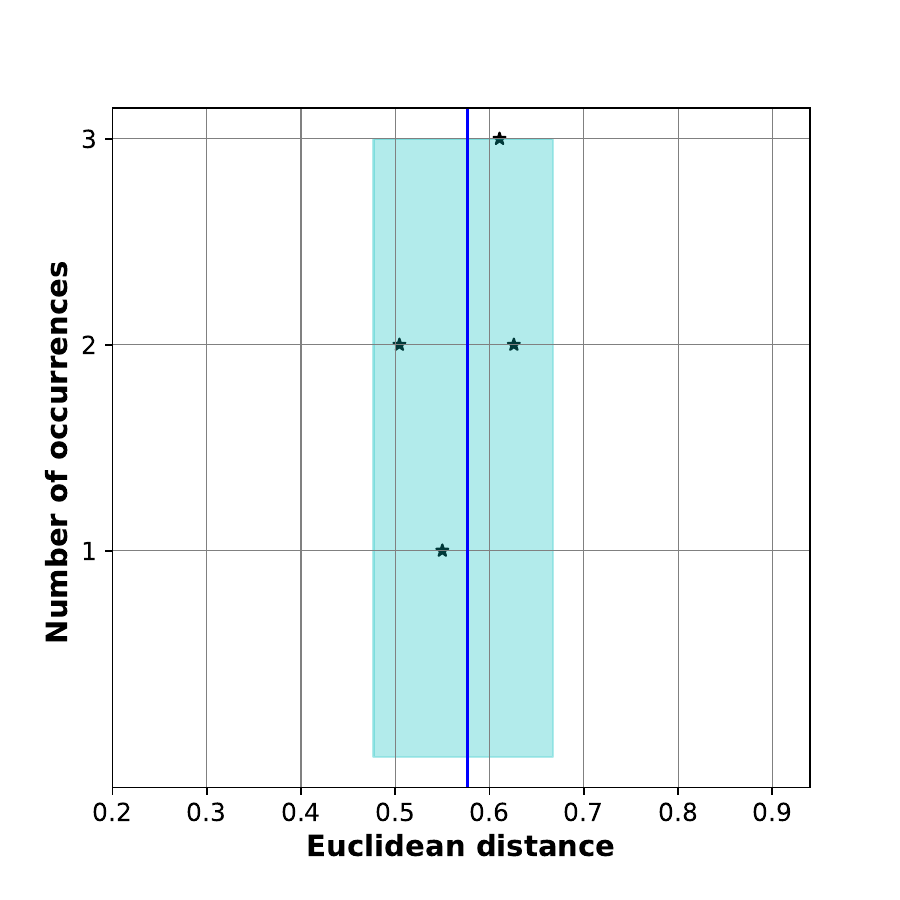}
    \caption{Random real-valued vectors (see Table~\ref{tab:rand} in
    Supp.~Mat.)}
    \label{fig:hist}
    \end{subfigure}
    \begin{subfigure}[b]{0.45\textwidth}
    \includegraphics[width=0.8\linewidth]{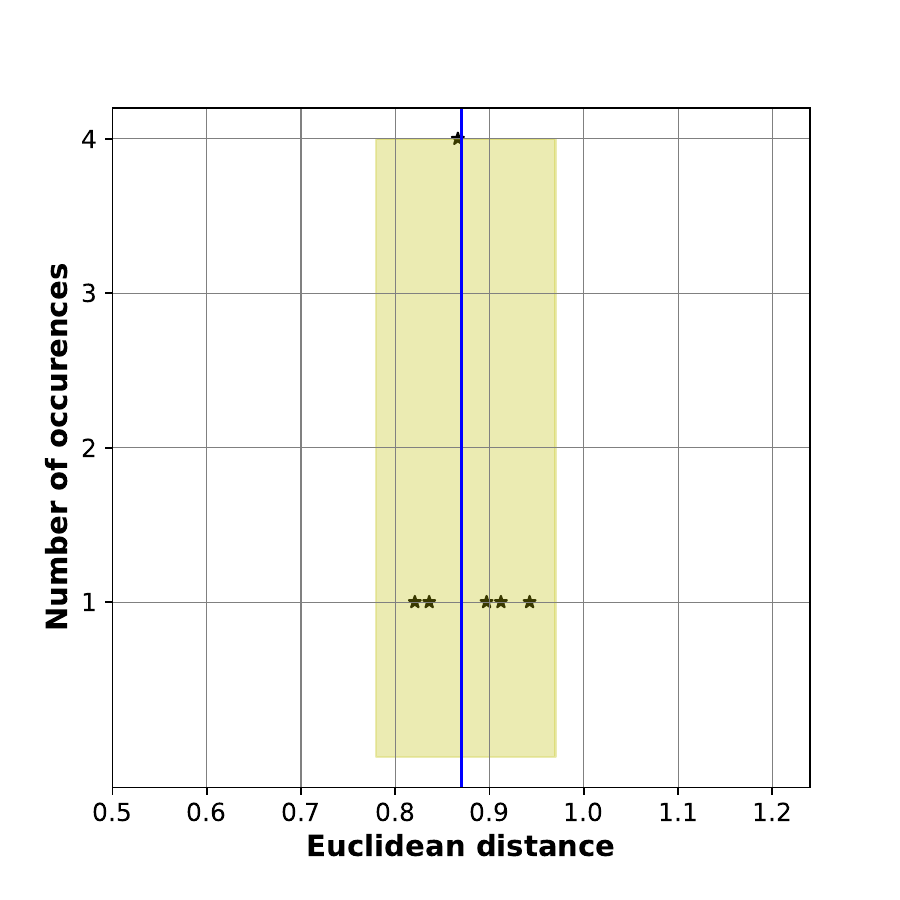}
    \caption{Grayscale images of Fig.~\ref{fig:moons} (see
    Table~\ref{tab:moons} in Supp.~Mat.)}
    \label{fig:hist2}
    \end{subfigure}
    \caption{Distributions of the estimated Euclidean distance over repeated
    protocol runs. The shaded pink region indicates the $\pm 0.1$ error
    bound.}
    \label{Fig:histograms}
\end{figure}

\vspace{0.5mm}
\noindent\textbf{Implementation.}
We first validated the setup using two boundary cases at input sizes
$n=10^7$ and $n=10^8$: identical data sets ($\tilde{E}=0$) and maximally
different data sets ($\tilde{E}=2$). We then tested the protocol on many random
real-valued vectors, one of which we show here with a known distance of ${E}=0.5769$. Finally,
we demonstrated the protocol on the grayscale images shown in
Fig.~\ref{fig:moons}, whose pixel values, mapped to integers from 0 to 9,
fall within our modulation range. The Euclidean distance between these images
is ${E}=0.8799$. For the grayscale data set at input size $n=10^8$, we
demonstrate a quantum advantage in transmitted information over the best known
classical protocol, as shown in Fig.~\ref{fig:EDplot}.

\vspace{0.5em}
 
\noindent\textbf{Phase fluctuation tracking.}
Real-time monitoring of the interferometric phase is essential
in the absence of an active phase-locking mechanism. The phase
is tracked continuously by sending equal-amplitude pulses on
both arms and monitoring the click statistics on $D_0$ and
$D_1$: when the inputs are equal ($x_j = y_j$ for all $j$),
detector $D_1$ ideally receives no light, so the visibility
$\mathcal{V} = (C_0 - C_1)/(C_0 + C_1)$ computed from the click
counts $C_0$ and $C_1$ provides a direct real-time estimate
of the interferometric phase stability. In our setup, the
phase typically remains stable over timescales of order one
second. Once the visibility exceeds the threshold $\mathcal{V}_{\min}$
established during pre-processing (Protocol~\ref{prot:pre},
item~6), the equal-amplitude monitoring pulses are replaced
by amplitude-modulated pulses encoding the actual input
vectors $X$ and $Y$, and data acquisition begins. The full
fingerprint of $n$ pulses is not sent in a single
uninterrupted burst; instead, the transmission is divided
into chunks of approximately $5 \times 10^3$ pulses, each
lasting one millisecond at the 5~MHz pulse repetition rate.
Chunks are transmitted in the correct order one by one only when the visibility is above
$\mathcal{V}_{\min}$; if the phase drifts between chunks and the
visibility drops below threshold, transmission is suspended
and monitoring resumes until stability is recovered. In this
way, the $n = 10^7$ and $n = 10^8$ input sizes correspond to
$2 \times 10^3$ and $ 2 \times 10^4$ such chunks respectively, accumulated over
multiple stable windows during the protocol run.
  
\vspace{0.5em}
\noindent\textbf{Post-processing.}\label{sec:postpro}

The Euclidean distance is estimated and validated following
Protocol~\ref{prot:post}.  
We verify that the estimated distance satisfies the additive
error bound $|\tilde{E} - \|x-y\|_2^2| \leq \epsilon = 0.1$
with failure probability $\delta = 10^{-6}$,   
consistent with the Chernoff bound of~\cite{Niraj}. Each experiment was
repeated between 6 and 10 times, and we checked whether any run produced an
estimate outside the $\pm 0.1$ interval around the true distance. The
resulting distributions of estimated distances are shown as histograms in
Figs.~\ref{fig:hist} and~\ref{fig:hist2}; detailed numerical results for each
run are provided in the Supplementary Material. In all cases, every run fell
within the required error bound.    To evaluate the quantum advantage quantitatively, we compute the
transmitted information at $n=10^8$ using the experimental mean photon number and detector quantum efficiency. The
number of repetitions of the classical protocol required to achieve
failure probability $\delta = 10^{-6}$ is $k = \lceil \log\delta /
\log(1/4) \rceil = 10$. This gives a total quantum transmitted
information of $I_Q \approx270{,}424$~bits, compared to the best
known classical protocol bound $I_C^{\mathrm{best}} = 4k\sqrt{n}
\approx 400{,}000$~bits, yielding a quantum advantage factor of
$\mathcal{A}_{\mathrm{best}} \approx 1.48$. We note that both
protocols are evaluated at the same failure probability $\delta =
10^{-6}$ and the same ED accuracy $\epsilon = 0.1$, ensuring a
fair comparison. The classical lower bound $I_C^{\mathrm{lb}}
\approx 13{,}589$~bits has not yet been beaten at this input size,
and would require larger $n$ or lower $\mu$ (see detailed calculations in Appendix~\ref{app:error}).    

\refstepcounter{protocol}\label{prot:pre}
\begin{framed}
\noindent\textbf{Pre-processing Protocol~\theprotocol: Calibration
and Selection of $\mu$}
\vspace{0.5em}

\noindent\textbf{Input:} Target ED accuracy $\epsilon$, failure
probability $\delta$, input size $n$, detector quantum
efficiencies $\eta_A$ and $\eta_B$ of $D_0$ and $D_1$
respectively.

\noindent\textbf{Goal:} Determine the mean photon number
$\mu^*$ to be sent by each party, accounting for detector
quantum efficiency, and verify interferometer visibility
$\nu$ before running the protocol.

\vspace{0.5em}
\begin{enumerate}

\item \textbf{Fix target precision.}
Choose the desired ED accuracy $\epsilon$ and failure
probability $\delta$. The protocol guarantees
$|\tilde{E} - \|x-y\|_2^2| \leq \epsilon$ with probability
at least $1-\delta$.

\item \textbf{Solve for the effective detected photon number
$\mu_{\mathrm{ref}}$.}
Since the single-photon detectors have quantum efficiencies
$\eta_{A}$ and $\eta_B$, only a fraction of the arriving
photons produce clicks. The error bound depends on the
number of detected photons $\mu_{\mathrm{ref}}$, not the
number sent. Find $\mu_{\mathrm{ref}}$ numerically by
solving $P_{\mathrm{error}}(\mu_{\mathrm{ref}}, n)
= \delta$, where the error probability under the Gaussian
approximation is:
\begin{equation*}
    P_{\mathrm{error}}(\mu_{\mathrm{ref}}, n) =
    1 - \Phi\!\left(\frac{\epsilon}{\sigma}\right)
    + \Phi\!\left(\frac{-\epsilon}{\sigma}\right),
\end{equation*}   
\begin{equation*}
    \sigma = \frac{\sqrt{2(\mu_{\mathrm{ref}}
    + n\,p_d)}}{\mu_{\mathrm{ref}}(2\nu-1)},
\end{equation*}
with $\Phi$ the standard normal CDF. This criterion is
tighter than the analytic Chernoff bound and is the one
used in practice (see Appendix~\ref{app:chernoff} for more details).

\item \textbf{Compute $\mu^*$ to be sent and set the VOAs.}
Since each detector registers only $\eta_{A,B}$
of the photons arriving at it, Alice and Bob must send
more photons than $\mu_{\mathrm{ref}}$ so that the
correct number of clicks is produced. The mean photon
number to be sent is:
\begin{equation*}
    \mu^*_A = \frac{\mu_{\mathrm{ref}}}{\eta_A},
    \qquad
    \mu^*_B = \frac{\mu_{\mathrm{ref}}}{\eta_B}.
\end{equation*}
The variable optical attenuators (VOAs) are adjusted so
that Alice and Bob send $\mu^*_A$ and $\mu^*_B$ mean
photons respectively over the full fingerprint of $n$
pulses. The mean photon number per pulse is:
\begin{equation*}
    \bar{\mu}^{A}_{\mathrm{pulse}} = \frac{\mu^*_A}{n}
    = \frac{\mu_{\mathrm{ref}}}{\eta_A\, n}, \qquad
    \bar{\mu}^{B}_{\mathrm{pulse}} = \frac{\mu^*_B}{n}
    = \frac{\mu_{\mathrm{ref}}}{\eta_B\, n}.
\end{equation*}

\item \textbf{Verify the quantum advantage condition.}
Compute the total quantum transmitted information $I_Q =
I_Q^A + I_Q^B$ (see Protocol~\ref{prot:post}, item~4)
and confirm:
\begin{equation*}
    I_Q \;<\; I_C^{\mathrm{best}} = 4k\sqrt{n},
    \qquad k = \left\lceil
    \frac{\ln\delta}{\ln(1/4)}\right\rceil.
\end{equation*}

\item \textbf{Calibrate $\mu^*$ via short-run measurement.}
Rather than waiting for the full $n$-pulse transmission,
send $n_{\mathrm{cal}} \ll n$ maximum-amplitude pulses
($x_j = y_j = 1$) and measure the total click count $R$
on either detector. The effective detected mean photon
number is inferred as:
\begin{equation*}
    \mu_{\mathrm{ref}} \approx \frac{R}{n_{\mathrm{cal}}},
\end{equation*}
valid in the weak-pulse regime. If the inferred
$\mu_{\mathrm{ref}}$ differs from the target value
found in item~2, the VOA is fine-adjusted to correct
$\mu^*_{A,B} = \mu_{\mathrm{ref}}/\eta_{A,B}$
accordingly.

\item \textbf{Measure and threshold visibility $\mathcal{V}$.}
Using the same calibration run, estimate the
interferometer visibility as:
\begin{equation*}
    \mathcal{V} = \frac{C_0 - C_1}{C_0 + C_1},
\end{equation*}
where $C_0$ and $C_1$ are click counts at $D_0$ and
$D_1$ with equal maximum-amplitude inputs ($x_j = y_j
= 1$, so $D_1$ ideally sees no light). Proceed only
if $\mathcal{V} \geq \mathcal{V}_{\min}$, where $\mathcal{V}_{\min}$ is
consistent with the Gaussian criterion of item~2 for
the target $(\epsilon, \delta)$.

\end{enumerate}

\noindent\textbf{Output:} Mean photon numbers $\mu^*_A$
and $\mu^*_B$ to be sent (VOA settings), effective
detected mean photon number $\mu_{\mathrm{ref}}$,
and calibrated visibility $\nu$.
\end{framed}

\refstepcounter{protocol}\label{prot:post}
\begin{framed}
\noindent\textbf{Post-processing Protocol~\theprotocol:
Distance Estimation and Advantage Verification}
\vspace{0.5em}

\noindent\textbf{Input:} Raw detector click records
$\{Z_j^{D_0}, Z_j^{D_1}\}_{j=1}^{n}$ over
$N_{\mathrm{runs}}$ repetitions, calibrated parameters
$(\mu_{\mathrm{ref}}, \mu^*_A, \mu^*_B, \nu, \eta_A,
\eta_B)$ from Protocol~\ref{prot:pre}, true distance
$\|x-y\|_2^2$ computed classically.

\noindent\textbf{Goal:} Estimate the Euclidean distance,
verify the error bound, and confirm the quantum advantage.

\vspace{0.5em}
\begin{enumerate}

\item \textbf{Compute the signed click sum.}
For each run $r \in \{1,\ldots,N_{\mathrm{runs}}\}$:
\begin{equation*}
    \Delta Z^{(r)} = \sum_{j=1}^{n}
    \Bigl(Z_j^{D_0} - Z_j^{D_1}\Bigr).
\end{equation*}
In expectation: $\mathbb{E}[\Delta Z] =
2\mu_{\mathrm{ref}}(2\nu-1)\langle x,y\rangle$,
where $\langle x,y\rangle = \sum_j x_j y_j$.
Dark-count contributions cancel out in the expectation.

\item \textbf{Estimate the Euclidean distance.}
For each run $r$:
\begin{equation*}
    \tilde{E}^{(r)} = 2 - \frac{\Delta Z^{(r)}}
    {\mu_{\mathrm{ref}}(2\nu - 1)},
\end{equation*}
where $\mu_{\mathrm{ref}}$ is the effective detected
mean photon number calibrated in
Protocol~\ref{prot:pre}. The estimator standard
deviation is:
\begin{equation*}
    \sigma = \frac{\sqrt{2(\mu_{\mathrm{ref}}
    + n\,p_d)}}{\mu_{\mathrm{ref}}(2\nu-1)}.
\end{equation*}

\item \textbf{Verify the error bound.}
For each run $r$, check:
\begin{equation*}
    \bigl|\tilde{E}^{(r)} - \|x - y\|_2^2\bigr|
    \leq \epsilon.
\end{equation*}
The empirical failure rate must satisfy:
\begin{equation*}
    \hat{\delta} = \frac{\#\{\text{failed runs}\}}
    {N_{\mathrm{runs}}} \lesssim \delta.
\end{equation*}

\item \textbf{Compute the quantum transmitted
information.}
The transmitted information accounts for the
Poissonian spread $\Delta N$ of the photon number
distribution and the detector quantum efficiencies.
Since each detector registers only $\eta_{A,B}$ of
the arriving photons, the referee effectively has
access to $\mu^*_{A,B} = \mu_{\mathrm{ref}}/
\eta_{A,B}$ photons sent per party, and $\Delta N$
is computed at this value. The spread $\Delta N$ is
the smallest integer satisfying:
\begin{multline*}
    \mathrm{Poi}\!\left(N \leq \mu^*_A
    - \Delta N;\,\mu^*_A\right) \\
    + \left[1 - \mathrm{Poi}\!\left(N \leq \mu^*_A
    + \Delta N;\,\mu^*_A\right)\right]
    \leq \epsilon_{\mathrm{td}},
\end{multline*}
and analogously for $\mu^*_B$, where
$\epsilon_{\mathrm{td}} = 10^{-6}$. The quantum
transmitted information per party is then:
\begin{equation*}
    I_Q^{A} = \log_2(2\Delta N_A) +
    \log_2\binom{n + \mu^*_A + \Delta N_A - 1}{n-1},
\end{equation*}

\begin{equation*}    
    I_Q^{B} = \log_2(2\Delta N_B) +
    \log_2\binom{n + \mu^*_B + \Delta N_B - 1}{n-1},
\end{equation*}
and the total is $I_Q = I_Q^A + I_Q^B$.
For each experimental run $r$, the transmitted information
$I_Q^{(r)}$ is computed from the calibrated mean photon number
$\mu^{(r)}$ measured at the start of that run. The experimental
points in Fig.~\ref{fig:EDplot} show the mean $\bar{I}_Q$ over
all runs, with error bars representing the standard deviation
across runs arising from shot-to-shot variations in $\mu^{(r)}$
within the operating range $\mu \in [5000, 7500]$.

\item \textbf{Confirm the quantum advantage.}
Compute the classical bounds~\cite{Niraj}:
\begin{equation*}
    I_C^{\mathrm{best}}(n) = 4k\sqrt{n},
    \quad k = \left\lceil\frac{\ln\delta}{\ln(1/4)}
    \right\rceil,
\end{equation*}

\begin{equation*}    
    I_C^{\mathrm{lb}}(n) = 2\sqrt{g_3(\epsilon)}
    \,\sqrt{n} - g_3(\epsilon),
\end{equation*}

\begin{equation*}
    g_3(\epsilon) = 2\log_2(e)
    \!\left(\tfrac{1}{2}-\epsilon\right)^2.
\end{equation*}
The advantage factors are:
\begin{equation*}
    \mathcal{A}_{\mathrm{best}} =
    \frac{I_C^{\mathrm{best}}}{I_Q}, \qquad
    \mathcal{A}_{\mathrm{lb}} =
    \frac{I_C^{\mathrm{lb}}}{I_Q}.
\end{equation*}
The advantage is demonstrated whenever
$\mathcal{A}_{\mathrm{best}} > 1$.

\item \textbf{Report results.}
Collect $\{\tilde{E}^{(r)}\}$, display as a
histogram, confirm all runs lie within $\pm\epsilon$
of the true distance, and report $I_Q$,
$I_C^{\mathrm{best}}$, $I_C^{\mathrm{lb}}$, and
$\mathcal{A}_{\mathrm{best}}$,
$\mathcal{A}_{\mathrm{lb}}$ for each $n$.

\end{enumerate}

\noindent\textbf{Output:} Estimated distances
$\{\tilde{E}^{(r)}\}$, empirical failure rate
$\hat{\delta}$, advantage factors
$\mathcal{A}_{\mathrm{best}}$ and
$\mathcal{A}_{\mathrm{lb}}$, and total quantum
transmitted information $I_Q$.
\end{framed}

\section{Discussion}
We have experimentally demonstrated an exponential quantum advantage in
transmitted information over the best known classical protocol for the
Euclidean distance problem in the simultaneous message passing model of
communication complexity, using real-valued data sets. The precision of the
distance estimator was confirmed to lie within the required error bounds for
both random vectors and grayscale images.

Several directions remain open for improving the performance of the
demonstrated protocol. Detectors with even lower dark-count rates would
directly reduce the optimal $\mu$, shrinking the transmitted information and
widening the region of quantum advantage. On the stability side, a
phase-locking feedforward mechanism would provide a more robust and systematic
approach to maintaining the interferometer visibility, replacing the
threshold-based gating currently used to handle phase fluctuations.

The present experiment is also limited in pulse rate to 5~MHz by the
electronics. Increasing this rate would allow access to larger input sizes,
further extending the demonstrated region of quantum advantage and potentially
enabling the protocol to beat the classical lower bound.

A complementary direction would be to recover the quantum advantage in
communication time, which is not demonstrated in the current implementation.
Multiplexing techniques, studied theoretically for the ED problem
in~\cite{Niraj} and demonstrated experimentally for the Equality problem
in~\cite{effprint}, offer a promising route. However, the multiplexing
scheme of~\cite{effprint} operates without demultiplexing, relying only on
total detector counts rather than per-pulse click records. Since the ED
estimator requires per-pulse resolution, adapting such a scheme to the ED
problem remains a non-trivial open question.

Our results show that quantum advantages in communication complexity are not
restricted to binary data processing tasks such as Equality, but extend to
richer, real-valued problems with direct relevance to practical applications
such as data similarity search and machine learning.\\

\noindent {\large \textbf{Acknowledgments}}\\
We acknowledge financial support from the European Research Council project QUSCO and the PEPR integrated project EPiQ (ANR-22-PETQ-0007), which is part of Plan France 2030.\\

\bibliographystyle{apsrev4-2}
\bibliography{biblio}

\appendix
\section{Data examples for \texorpdfstring{$n=10^8$}{n=108}}
       
\begin{table}[h!]
\caption{The experimental counts and calculated ED of a random vector with a distance $\Tilde{E}=0.5769$. Also represented in Fig. \ref{fig:hist} }
\label{tab:rand}
\begin{ruledtabular}
\begin{tabular}{cccc}
\hline
Counts $D_0$ & Counts $D_1$ & Difference & ED \\
\hline
5729.0 & 631.0 & 5098.0 & 0.617978 \\ 
6724.0 & 743.0 & 5981.0 & 0.618980 \\ 
7123.0 & 643.0 & 6480.0 & 0.561368 \\ 
6063.0 & 718.0 & 5345.0 & 0.640979 \\ 
6427.0 & 707.0 & 5720.0 & 0.617596 \\ 
6473.0 & 902.0 & 5571.0 & 0.489220 \\ 
5896.0 & 1094.0 & 4802.0 & 0.626037 \\ 
7201.0 & 1066.0 & 6135.0 & 0.515785 \\ 
6962.0 & 1028.0 & 5934.0 & 0.514643 \\ 
\hline
\end{tabular}
\end{ruledtabular}
\end{table}

%

\begin{table}[h!]
\caption{The experimental counts and calculated ED of the grayscale images with a distance $\Tilde{E}=0.8799$. Also represented in Fig. \ref{fig:hist2} }
\label{tab:moons}
\begin{ruledtabular}
\begin{tabular}{cccc}
\hline
Counts $D_0$ & Counts $D_1$ & Difference & ED \\
\hline
 4697.0 & 582.0 & 4115.0 & 0.917355 \\ 
 5244.0 & 550.0 & 4694.0 & 0.874793 \\ 
 5118.0 & 587.0 &4531.0 & 0.896922 \\ 
 5206.0 &440.0 & 4766.0 & 0.827586 \\
 5632.0 & 522.0 & 5110.0 & 0.846730 \\
 5886.0 & 443.0 &5443.0 & 0.805542 \\
 5886.0 &606.0 & 5280.0 & 0.870404 \\
 5148.0 & 734.0 &4414.0 & 0.957743 \\
 5528.0 &580.0 & 4948.0 & 0.874881 \\
 5960.0 & 633.0 & 5327.0 & 0.877808 \\
\hline
\end{tabular}
\end{ruledtabular}
\end{table}

\section{Supplementary Derivation: Chernoff Bound and
Optimal \texorpdfstring{$\mu^*$}{mu*}} \label{app:chernoff}

We compare here the form of the Chernoff bound as it appears in
the three relevant references~\cite{Niraj, Ar, Xu} and identify
precisely where the inconsistency in our Eq.~(\ref{eq:chernoff})
originates.

\subsection*{Standard Chernoff bound}

The standard multiplicative Chernoff bound for a sum
$S = \sum_i X_i$ of independent Bernoulli random variables
with mean $\mathbb{E}[S] = M$ reads~\cite{mitzenmacher2017}:
\begin{equation}
    \Pr(|S - M| \geq \epsilon M) \leq
    2\left(\frac{e^{\epsilon}}{(1+\epsilon)^{1+\epsilon}}
    \right)^{M}.
    \label{eq:chernoff_standard}
\end{equation}
This is the form cited throughout the fingerprinting
literature and is valid whenever $\epsilon > 0$ and the
$X_i$ are independent.

\subsection*{Form used in Xu et al.\ \texorpdfstring{\cite{Xu}}{[Xu]}}

Xu et al.\ implement the Arrazola protocol experimentally
for binary (Equality) data. Their Chernoff analysis
follows~\cite{Ar} directly, applying the standard bound
with $M = \mu$ at the ideal visibility ($\nu = 1$).
Visibility imperfections are accounted for separately
through an effective mean photon number but do not modify
the structure of the Chernoff exponent. Their bound is
therefore also the standard form at $r = 1$:
\begin{equation}
    \delta = 2\left(\frac{e^{\epsilon}}
    {(1+\epsilon)^{1+\epsilon}}\right)^{\mu}.
\end{equation}
Neither~\cite{Ar} nor~\cite{Xu} encounter the issue
described below because they never use the $D_0 - D_1$
difference estimator that introduces the factor $r < 1$
when $\nu < 1$.

\subsection*{Form in Kumar et al.} Kumar et al.~\cite{Niraj}

Kumar et al. introduce the $D_0 - D_1$ difference
estimator for the Euclidean distance problem, whose
expected value is $\mathbb{E}[\Delta Z] = 2\mu(2\nu-1)
\langle x,y\rangle$. The effective fluctuation parameter
entering the Chernoff bound is no longer $\epsilon$ but
$r\epsilon$, where:
\begin{equation}
    r = \frac{2\nu-1}{\nu +
    \dfrac{n-2\mu}{2\mu}\,p_d} \leq 1,
\end{equation}
with $r = 1$ only at $\nu = 1$, $p_d = 0$. Substituting
$\epsilon \to r\epsilon$ into
Eq.~(\ref{eq:chernoff_standard}) yields the correct form:
\begin{equation}
    \delta = 2\left(\frac{e^{r\epsilon}}
    {(1+r\epsilon)^{1+r\epsilon}}\right)^{2\mu(2\nu-1)}.
    \label{eq:chernoff_correct}
\end{equation}
Since the detectors have quantum efficiencies $\eta_A$
and $\eta_B$, the mean number of photons that must be
\emph{sent} by Alice and Bob to achieve
$\mu_{\mathrm{ref}}$ clicks is:
\begin{equation}
    \mu^*_A = \frac{\mu_{\mathrm{ref}}}{\eta_A},
    \qquad
    \mu^*_B = \frac{\mu_{\mathrm{ref}}}{\eta_B},
\end{equation}
and the VOAs are set to these values. We define the
log-bracket:
\begin{equation}
    L \equiv r\epsilon - (1+r\epsilon)\ln(1+r\epsilon).
\end{equation}

\subsection*{Checking that \texorpdfstring{$L < 0$}{L < 0} always }

For any $x > 0$, $f(x) = x - (1+x)\ln(1+x)$ satisfies
$f(0)=0$ and $f'(x) = -\ln(1+x) < 0$, so $f(x) < 0$
for all $x > 0$. With $x = r\epsilon > 0$:
\begin{equation}
    L = r\epsilon - (1+r\epsilon)\ln(1+r\epsilon) < 0
    \quad \forall\; r,\epsilon > 0. \quad 
\end{equation}
The bound therefore decreases with $\mu_{\mathrm{ref}}$
for any $\nu > 1/2$, and can always be solved for
$\mu_{\mathrm{ref}}$.

\subsection*{ Asymptotic scaling of \texorpdfstring{$\mu_{\mathrm{ref}}$}{mu\_ref}}

Expanding $L$ for small $r\epsilon$ using
$\ln(1+x) \approx x - x^2/2$:
\begin{equation}
    L \approx -\frac{(r\epsilon)^2}{2}
    = -\frac{r^2\epsilon^2}{2}.
\end{equation}
The bound becomes:
\begin{equation}
    \delta \approx 2\exp\!\left(
    -\mu_{\mathrm{ref}}(2\nu-1)\,r^2\epsilon^2\right).
\end{equation}
Setting equal to $\delta$ and solving:
\begin{equation}
    \mu_{\mathrm{ref}} = \frac{\ln(2/\delta)}
    {(2\nu-1)\,r^2\,\epsilon^2}.
\end{equation}
In the regime $p_d \ll \mu_{\mathrm{ref}}/n$:
$r \approx (2\nu-1)/\nu$, giving:
\begin{equation}
    \mu_{\mathrm{ref}} =
    \frac{\nu^2\,\ln(2/\delta)}
    {(2\nu-1)^3\,\epsilon^2}.
    \label{eq:mu_star_asymptotic}
\end{equation}
Numerically at $\mathcal{V}=0.9$, $\epsilon=0.1$,
$\delta=10^{-6}$:
\begin{align}
    \mu_{\mathrm{ref}} &=
    \frac{0.9\times\ln(2\times10^6)}
    {0.73\times 0.01} \approx 1990.
\end{align}
The photons to be sent by each party (VOA setting) are
then:
\begin{equation}
    \mu^*_A = \frac{\mu_{\mathrm{ref}}}{\eta_A}
    = \frac{1990}{0.84} \approx 2369, \quad
    \mu^*_B = \frac{\mu_{\mathrm{ref}}}{\eta_B}
    = \frac{1990}{0.85} \approx 2341.
\end{equation}

\subsection*{Gaussian criterion}

Since $\Delta Z = \sum_{j=1}^n (Z_j^{D_0} - Z_j^{D_1})$ is a
sum of $n$ independent bounded random variables, the central
limit theorem guarantees that for large $n$ it is approximately
Gaussian with variance $\mathrm{Var}(\Delta Z) \approx
2(\mu_{\mathrm{ref}} + n\,p_d)$, where the two terms account
for signal clicks and dark counts respectively (the approximation
uses $p_j \ll 1$ in the weak-pulse regime). The probability that
the estimator $\tilde{E}$ deviates from the true distance by more
than $\epsilon$ is therefore:
\begin{equation}
    P_{\mathrm{error}} = 1 -
    \Phi\!\left(\frac{\epsilon}{\sigma}\right)
    + \Phi\!\left(\frac{-\epsilon}{\sigma}\right),
    \quad
    \sigma = \frac{\sqrt{2(\mu_{\mathrm{ref}} + n\,p_d)}}
    {\mu_{\mathrm{ref}}(2\nu-1)},
\end{equation}
where $\Phi$ is the standard normal CDF. A similar Gaussian
approximation is used in the experimental analysis
of~\cite{Xu}.\\
Solving numerically with $\nu=0.95$ (from
$\mathcal{V}=0.9$), $p_d=1.6\times10^{-7}$,
$\epsilon=0.1$, $\delta=10^{-6}$, $n=10^8$:
\begin{equation}
    \mu_{\mathrm{ref,Gaussian}} \approx 4791
\end{equation}
The VOA settings are then:
\begin{equation}
    \mu^*_A = \frac{4791}{0.84} \approx 5703,
    \qquad
    \mu^*_B = \frac{4791}{0.85} \approx 5636,
\end{equation}
both consistent with the experimental operating range
$\mu \in [5000, 7500]$, which provides a safety margin
against phase fluctuations and visibility degradation. We have to notice that the Gaussian approximation share the same asymptotic value of $\mu_{\mathrm{ref}}$, but it gives a higher value since the dark counts are also affecting the variance which is taken into account.

\section{Error calculation for the experimental points}
\label{app:error}

The quantum transmitted information for each experimental point
is computed from the mean photon number $\mu$ measured during
calibration, the detector quantum efficiencies $\eta_A = 0.84$
and $\eta_B = 0.85$, and the input size $n$. We detail each
step below.

\subsection*{ Mean photons sent (VOA setting)}
Alice and Bob must send more photons than
$\mu_{\mathrm{ref}}$ to produce $\mu_{\mathrm{ref}}$ detected clicks on average. Note that for this section we assume $\mu_{\mathrm{ref}}$ with worse estimation of experimental parameters, this is why it is much higher that the previous sections). The mean
photon number sent by each party is:
\begin{equation}
    \mu^*_A = \frac{\mu}{\eta_A}, \qquad
    \mu^*_B = \frac{\mu}{\eta_B}.
\end{equation}

\noindent\textbf{At $n=10^7$, $\mu=7100$:}
\begin{equation*}
    \mu^*_A = \frac{7100}{0.84} = 8452,
    \qquad
    \mu^*_B = \frac{7100}{0.85} = 8353.
\end{equation*}

\noindent\textbf{At $n=10^8$, $\mu=7300$:}
\begin{equation*}
    \mu^*_A = \frac{7300}{0.84} = 8690,
    \qquad
    \mu^*_B = \frac{7300}{0.85} = 8588.
\end{equation*}

\subsection*{ Poisson spread \texorpdfstring{$\Delta N$}{Delta N}}

The Poisson spread $\Delta N$ is the smallest integer such that
the tail probability of the photon number distribution satisfies:
\begin{widetext}
\begin{equation}
    \mathrm{Poi}(N \leq \mu_{\mathrm{sent}} - \Delta N;\,
    \mu_{\mathrm{sent}})
    + \left[1 - \mathrm{Poi}(N \leq \mu_{\mathrm{sent}}
    + \Delta N;\, \mu_{\mathrm{sent}})\right]
    \leq \epsilon_{\mathrm{td}} = 10^{-6}.
\end{equation}
\end{widetext}
This bounds the Hilbert-space dimension of the fingerprint to
$2\Delta N$ with trace-distance precision $\epsilon_{\mathrm{td}}$.
Solved numerically:

\noindent\textbf{At $n=10^7$, $\mu=7100$:}
\begin{equation*}
    \Delta N_A(\mu^*_A=8452) = 450,
    \qquad
    \Delta N_B(\mu^*_B=8353) = 447.
\end{equation*}

\noindent\textbf{At $n=10^8$, $\mu=7300$:}
\begin{equation*}
    \Delta N_A(\mu^*_A=8690) = 456,
    \qquad
    \Delta N_B(\mu^*_B=8588) = 454.
\end{equation*}

\subsection*{ Quantum transmitted information per party}

The transmitted information for each party is:
\begin{equation}
    I_Q^{A,B} = \log_2(2\Delta N_{A,B}) +
    \log_2\binom{n + \mu^*_{A,B}
    + \Delta N_{A,B} - 1}{n - 1}.
\end{equation}
The first term accounts for the spread in photon number;
the second counts the number of distinguishable
fingerprint states in a Fock space of $n$ modes with at
most $\mu_{\mathrm{sent}} + \Delta N$ photons.

\noindent\textbf{At $n=10^7$, $\mu=7100$:}
\begin{align*}
    I_Q^A &= \log_2(2 \times 450) +
    \log_2\binom{10^7 + 8452 + 450 - 1}{10^7 - 1} \\
          &= \log_2(900) + \log_2\binom{10{,}008{,}901}{9{,}999{,}999} \\
          &= 9.81 + 103{,}054 = 103{,}064 \text{ bits}, \\
    I_Q^B &= \log_2(894) + \log_2\binom{10{,}008{,}799}{9{,}999{,}999} \\
          &= 9.80 + 102{,}017 = 102{,}027 \text{ bits}, \\
    I_Q   &= I_Q^A + I_Q^B = 205{,}090 \text{ bits}.
\end{align*}

\noindent\textbf{At $n=10^8$, $\mu=7300$:}
\begin{align*}
    I_Q^A &= \log_2(912) +
    \log_2\binom{100{,}009{,}146}{99{,}999{,}999} \\
          &= 9.83 + 135{,}904 = 135{,}914 \text{ bits}, \\
    I_Q^B &= \log_2(908) +
    \log_2\binom{100{,}009{,}041}{99{,}999{,}999} \\
          &= 9.83 + 134{,}501 = 134{,}510 \text{ bits}, \\
    I_Q   &= I_Q^A + I_Q^B = 270{,}424 \text{ bits}.
\end{align*}

\subsection*{ Classical comparison}

The number of repetitions of the classical Equality protocol
required to achieve failure probability $\delta = 10^{-6}$ is:
\begin{equation}
    k = \left\lceil \frac{\ln \delta}{\ln(1/4)} \right\rceil
    = \left\lceil \frac{\ln 10^{-6}}{\ln(1/4)} \right\rceil
    = \left\lceil \frac{-13.816}{-1.386} \right\rceil
    = \lceil 9.97 \rceil = 10.
\end{equation}
The best known classical protocol bound is $I_C^{\mathrm{best}}
= 4k\sqrt{n}$:
\begin{align*}
    n=10^7: \quad I_C^{\mathrm{best}} &= 4 \times 10 \times
    \sqrt{10^7} = 126{,}491 \text{ bits}, \\
    n=10^8: \quad I_C^{\mathrm{best}} &= 4 \times 10 \times
    \sqrt{10^8} = 400{,}000 \text{ bits}.
\end{align*}
The quantum advantage factors are:
\begin{align*}
    n=10^7: \quad \mathcal{A} &= \frac{126{,}491}{205{,}090}
    = 0.617 \quad \text{(no advantage yet)}, \\
    n=10^8: \quad \mathcal{A} &= \frac{400{,}000}{270{,}424}
    = 1.479 \quad \text{(advantage demonstrated)}.
\end{align*}


\subsection*{Classical lower bound}

The classical lower bound on the transmitted information
for the ED problem in the private-coin SMP model is
inherited from the Equality problem via a reduction:
any classical protocol estimating $\|x-y\|_2^2$ within
additive accuracy $\epsilon$ can decide Equality as a
special case, and hence its communication cost is at
least that of the best classical Equality protocol.
Following~\cite{Niraj}, the lower bound on the transmitted
information is:
\begin{equation}
    I_C^{\mathrm{lb}}(n) = 2\sqrt{g_3(\epsilon)}\,\sqrt{n}
    - g_3(\epsilon),
    \qquad
    g_3(\epsilon) = 2\log_2(e)\!\left(\tfrac{1}{2}
    - \epsilon\right)^2,
    \label{eq:classical_lb}
\end{equation}
where $\epsilon = 0.1$ is the additive ED accuracy.
This bound applies in the private-coin SMP model and
is derived by reduction from the $\Omega(\sqrt{n})$
lower bound on the Equality problem~\cite{NewmanSzegedy}.
For the parameters of our experiment ($\epsilon=0.1$,
$\delta=10^{-6}$):
\begin{equation*}
    g_3(0.1) = 2\log_2(e)(0.4)^2 \approx 0.4621,
\end{equation*}
giving:
\begin{align*}
    n=10^7: \quad I_C^{\mathrm{lb}} &=
    2\sqrt{0.4621}\times\sqrt{10^7} - 0.4621
    \approx 4299 \text{ bits},\\
    n=10^8: \quad I_C^{\mathrm{lb}} &=
    2\sqrt{0.4621}\times\sqrt{10^8} - 0.4621
    \approx 13596  \text{ bits}.
\end{align*}
As noted in the main text, this lower bound has not yet
been beaten at $n=10^8$, where our experimental quantum
transmitted information $I_Q \approx 270{,}000$~bits
exceeds $I_C^{\mathrm{lb}} \approx 43{,}013$~bits.
Beating the lower bound would require either increasing
$n$ beyond $10^{10}$ or reducing $\mu$ below the values
achievable with our current detector technology.

\subsection*{ Error bars from \texorpdfstring{$\mu$}{mu} uncertainty}

The experimental $\mu$ lies in the range $[5000, 7500]$
due to shot-to-shot calibration variation. The transmitted
information at the extremes of this range gives the
asymmetric error bars shown in Fig.~\ref{fig:EDplot}:

\begin{center}
\begin{tabular}{lcccc}
\hline
 & $I_Q(\mu=6000)$ & $I_Q(\mu_{\mathrm{exp}})$
 & $I_Q(\mu=7500)$ & Spread \\
\hline
$n=10^7$ & $177{,}641$ & $205{,}090$ & $214{,}914$ & $18\%$ \\
$n=10^8$ & $227{,}551$ & $270{,}424$ & $276{,}940$ & $18\%$ \\
\hline
\end{tabular}
\end{center}

Note that even at $\mu_{\min}=6000$ and $n=10^8$, one has
$I_Q = 227{,}551 < I_C^{\mathrm{best}} = 400{,}000$,
confirming that the quantum advantage at $n=10^8$ is
robust across the full experimental operating range.

\end{document}